\documentclass{article}

    \usepackage[final]{neurips_2023}

\usepackage[utf8]{inputenc} 
\usepackage[T1]{fontenc}    
\usepackage{hyperref}       
\usepackage{url}            
\usepackage{booktabs}       
\usepackage{amsfonts}       
\usepackage{nicefrac}       
\usepackage{microtype}      
\usepackage{xcolor}         
\usepackage{graphicx} 

\usepackage{array}

\title{Precipitation Prediction Using an Ensemble of Lightweight Learners}

\author{%
Xinzhe ~Li \\
BDIL, Alibaba Cloud \\
Hangzhou, China \\
\texttt{xinzhe.lxz@alibaba-inc.com} \\
\And
Rui ~Sun \\
BDIL, Alibaba Cloud \\
Hangzhou, China \\
\texttt{shiyi.sr@alibaba-inc.com} \\
\And
Yiming ~Niu\thanks{as an intern} \\
BDIL, Alibaba Cloud \\
Hangzhou, China \\
\texttt{nym424155@alibaba-inc.com} \\
\And
Yao ~Liu \\
BDIL, Alibaba Cloud \\
Hangzhou, China \\
\texttt{john.ly@alibaba-inc.com} \\
}

\begin{document}

\maketitle

\begin{abstract}
Precipitation prediction plays a crucial role in modern agriculture and industry. However, it poses significant challenges due to the diverse patterns and dynamics in time and space, as well as the scarcity of high precipitation events.
To address this challenge, we propose an ensemble learning framework that leverages multiple learners to capture the diverse patterns of precipitation distribution. Specifically, the framework consists of a precipitation predictor with multiple lightweight heads (learners) and a controller that combines the outputs from these heads. The learners and the controller are separately optimized with a proposed 3-stage training scheme.
By utilizing provided satellite images, the proposed approach can effectively model the intricate rainfall patterns, especially for high precipitation events. It achieved 1st place on the core test as well as the nowcasting leaderboards of the Weather4Cast 2023 competition. For detailed implementation, please refer to our GitHub repository at: https://github.com/lxz1217/weather4cast-2023-lxz
\end{abstract}

\section{Introduction}

Weather forecasting plays an important role in many applications of modern agriculture and industry such as efficient resource allocation, risk management, and decision-making processes. In recent years, Deep Neural Networks (DNNs) based methods have been widely used in this area due to their powerful learning ability. For example, Pangu-Weather~\cite{panguweather} and FourCastNet~\cite{fourcastnet} can effectively model global climate variables with a large-scale reanalysis dataset~\cite{era52, era51}.

However, precipitation prediction task is still challenging for DNNs due to two main factors that significantly impact its accuracy. 
Firstly, the sparsity of high precipitation events poses a significant obstacle, making it difficult for researchers to collect enough training data for DNNs. The limited availability of training samples hinders deep predictive networks to accurately model the spatial distribution as well as rainfall intensity of these infrequent events.
Secondly, the diverse and intricate patterns observed across different time steps, locations, and precipitation levels further hinder accurate prediction. Rainfall patterns often vary widely, exhibiting spatial heterogeneity and temporal variability. The intricate interplay between atmospheric conditions, geographical features, and local climate dynamics contributes to the complex nature of precipitation. Capturing and understanding the diverse patterns and dynamics becomes essential to develop effective models.

To tackle above challenges, in this paper, we propose an ensemble learning framework with lightweight learners for precipitation prediction. 

Specifically, WeatherFusionNet~\cite{weatherfusionnet} is adopted as the backbone network that gathers the information from its physical-informed and data-driven components, and we re-train its final U-Net component along with an additional ConvLSTM~\cite{convlstm} module. This combination enables our model to effectively tackle the long-term temporal correlations in the precipitation data.
Moreover, an ensemble of multiple lightweight heads (learners with one convolution layer) is exploit as the final prediction module. Multiple learners allow the framework to effectively capture the diverse patterns of precipitation distribution.
Like many MoE-based methods~\cite{moe1, moe2}, we introduce a controller model to generate weight maps for learners to make more reliable decisions. These weight maps serve to concentrate the learners’ attention on regions with a higher likelihood of rainfall occurring and disregarding regions without rainfall.
The model parameters of our framework are optimized with a proposed 3-stage training scheme that ensuring each module can be trained sufficiently during each training stage.

The proposed precipitation prediction approach achieves satisfying performance on Weather4cast 2023 Competition. Specifically, it won 1st place on the challenging core test and nowcasting leaderboards, 2nd place on the transfer-learning leaderboard. 

The rest of this paper is organized as follows: we first introduce each important component of the framework in Section 2. The proposed 3-stage training scheme is introduced in Section 3. Detailed experimental results are provided in Section 4.

\section{Methods}

Our proposed framework utilizes the WeatherFusionNet\cite{weatherfusionnet} as the backbone, as shown in the left part of Fig~\ref{fig:overall}. It consists of three main components: 

\begin{figure}[t]
    \centering
    \includegraphics[width=1\linewidth]{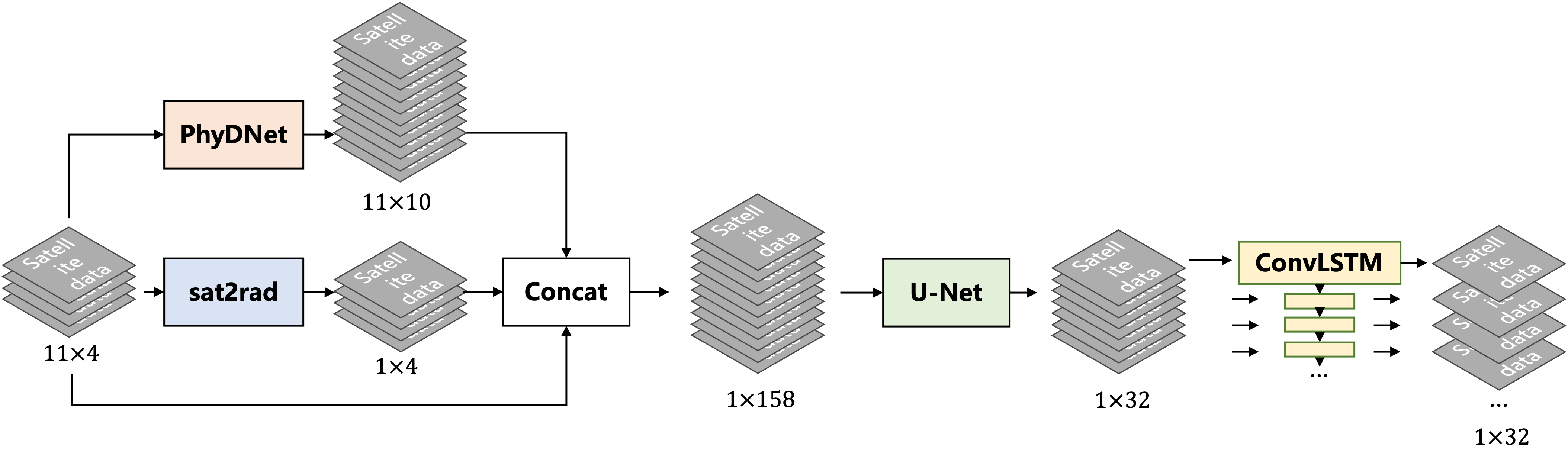}
    \caption{The proposed baseline network structure "WeatherFusionNet + ConvLSTM". The baseline contains 4 main components: PhyDNet, sat2radnet, U-Net as well as ConvLSTM. The parameters of first two networks are frozen during all the training process. }
    \label{fig:overall}
\end{figure}

\begin{itemize}
\item [1)]
\textbf{PhyDNet}~\cite{phydnet} is trained to predict the future frames of satellite data, enabling to model the temporal evolution of the weather patterns. PhyDNet disentangles physical dynamics from other visual information. 
\item [2)]
\textbf{sat2rad} network is trained to inference the precipitation levels of the current time frame. By training with such objective, sat2rad are expected to provide the modeling ability of extracting precipitation information from the satellite data. 
\item [3)]
\textbf{2D U-Net} is further employed to predict precipitation using the outputs from the two previous modules as well as the original satellite data. To be specific, the PhyDNet output consists of 11 channels across 10 frames, the sat2rad output contains 1 channel across 4 frames, and the original satellite data consists of 11 channels across 4 frames. To prepare the input data, the temporal dimension is flattened and the data is concatenated along the channel dimension, resulting in a total of 158 channels. The U-Net is implemented with 158 input channels and 32 output channels.
\end{itemize}

U-Net can’t model temporal information for the predicted frames due to its 2D network structure. We propose to improve the output with a ConvLSTM module, which processes the input frames step-by-step and enhance them with generated hidden states at each time step. 

\begin{figure}[t]
    \centering
    \includegraphics[width=0.60\linewidth]{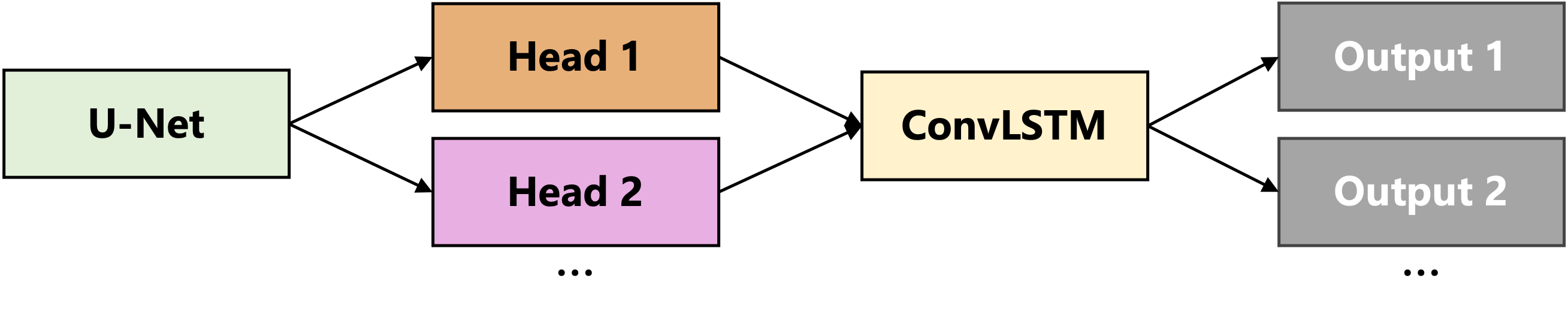}
    \caption{multiple output heads is attached to the U-Net, the ConvLSTM is shared for all heads}
    \label{fig:enter-label}
\end{figure}

To tackle the challenge posed by the diverse precipitation patterns in the training data, we propose an ensemble approach using multiple lightweight learners to improve the predictions. 
Our observation reveals that the last convolution layer of the U-Net transforms the hidden representation to the actual precipitation value, which can serve as a learner in the task of precipitation prediction. 
To leverage this insight, we integrate multiple such output heads and train each one independently with different settings (see Fig~\ref{fig:enter-label}). 

\begin{figure}[t]
    \centering
    \includegraphics[width=1\linewidth]{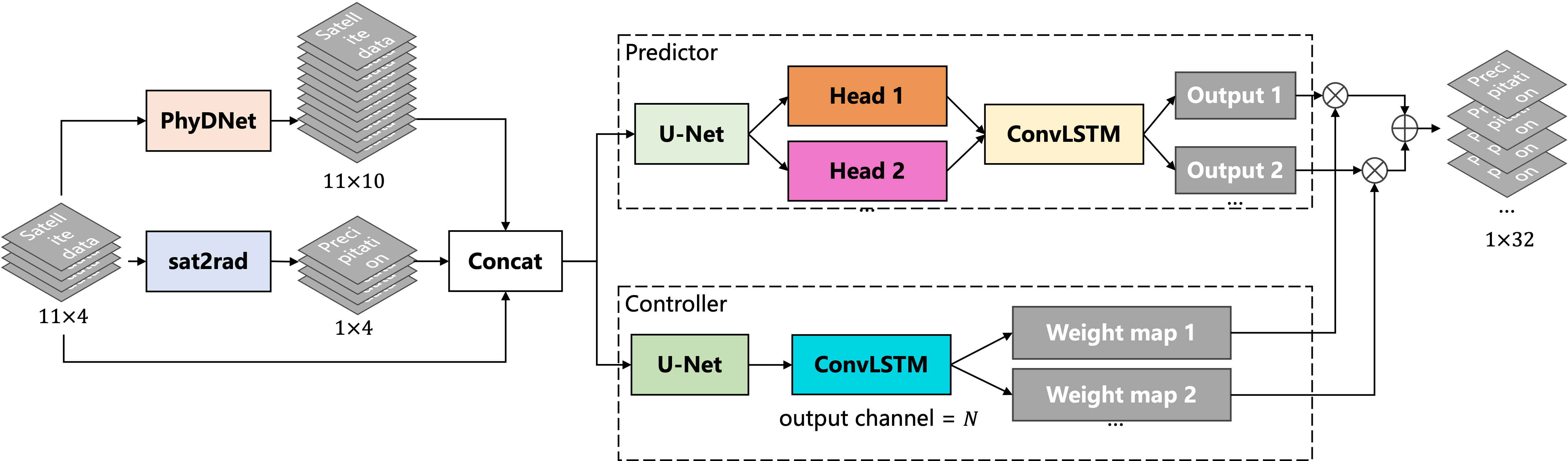}
    \caption{The controller is implemented as a gate to control the output}
    \label{fig:withcontroller}
\end{figure}

The outputs of these learners are combined by using a controller network for obtaining more reliable results. 
It takes a rainfall probability map as input and generates weight maps for the learners. These generated weight maps are multiplied with the output values of their corresponding learners, and the resulting values are further summed to produce the final prediction, the whole process is illustrated in Fig~\ref{fig:withcontroller}. In our implementation, we use a one-layer ConvLSTM module as the controller and the rainfall probability map can be obtained from the off-the-shelf method~\cite{weatherfusionnet}. During the inference phase, this rainfall probability map can be further used as a mask to exclude regions that are predicted to have a low likelihood of rainfall. 

We aim to leverage the strengths of each individual module, effectively extracting and fusing information from multiple dimensions of the satellite data to improve the accuracy and reliability of precipitation predictions. The experimental results will demonstrate that the proposed method has led to a substantial improvement in prediction accuracy, particularly in scenarios with high precipitation levels.

\section{Framework Training}

Our framework contain many network modules and it's quite difficult to optimize all their parameters sufficiently in a joint manner, so we propose to use a 3-stage training scheme instead.


\subsection{Stage 1: Backbone Training}
\begin{figure}[t]
    \centering
    \includegraphics[width=0.75\linewidth]{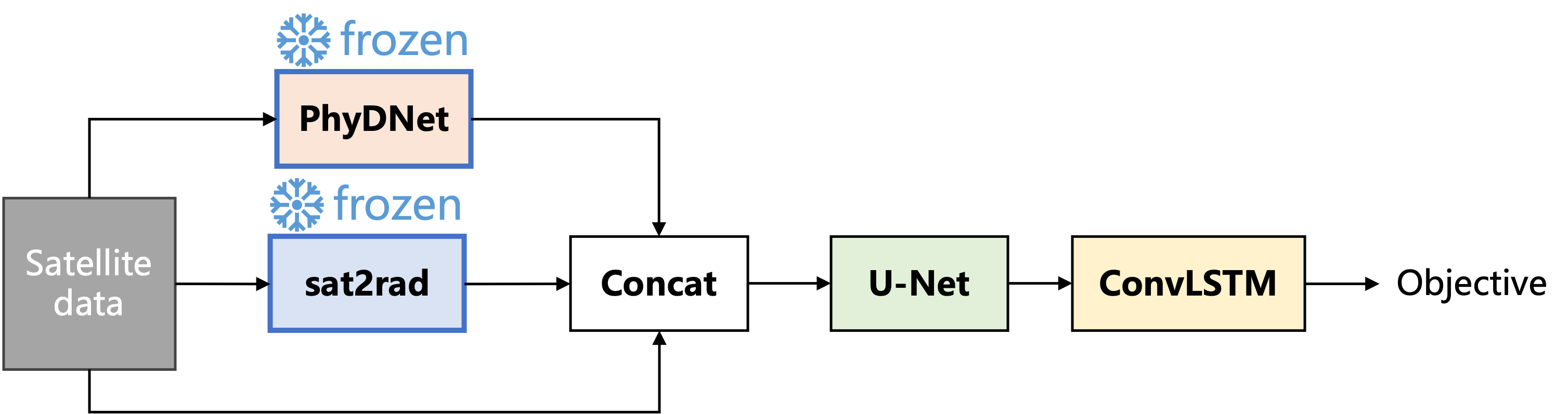}
    \caption{Backbone Training Stage}
    \label{fig:phrase2}
\end{figure}

The corresponding training process is illustrated in Fig~\ref{fig:phrase2}. 
To accurately predict future precipitation values, we re-train the U-Net of WeatherFusionNet, which utilizes the outputs of PhyDNet, sat2rad, and the original satellite data, along with a ConvLSTM module to model the temporal dependencies of precipitation values, thereby further enhancing the predictive performance. In this stage, the PhyDNet and sat2rad modules are frozen with pretrained parameters.

\subsection{Stage 2: Training of Ensemble Learners}
\begin{figure}[t]
    \centering
    \includegraphics[width=1\linewidth]{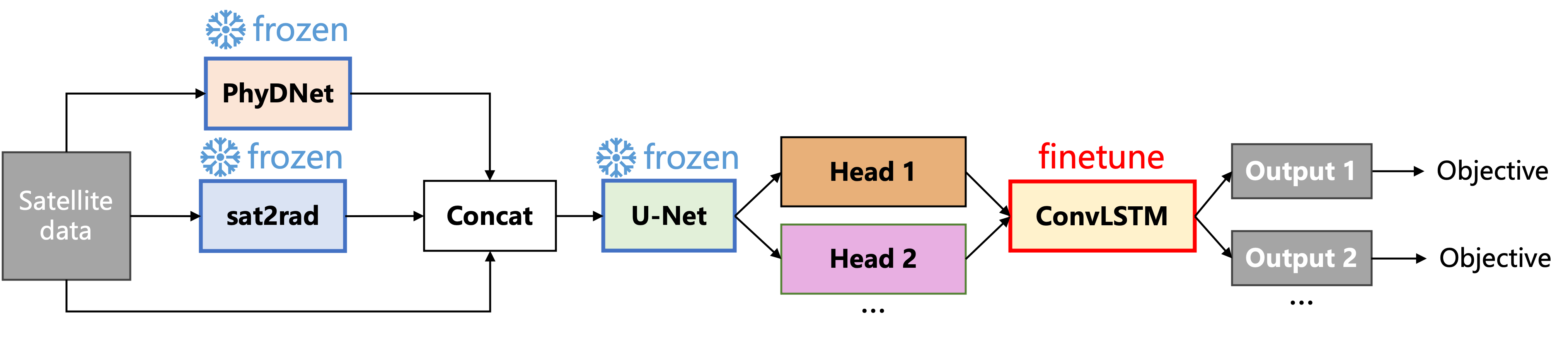}
    \caption{Training of Ensemble Learners}
    \label{fig:phase3}
\end{figure}

The corresponding training process is illustrated in Fig~\ref{fig:phase3}. 
The last convolution layer of the U-Net, referred to as the output head, plays the most important role in transforming the latent representation into actual precipitation values. The modeling capability and diversity of this layer in capturing the patterns within the latent representation directly impact the accuracy of the model's predictions. In this stage, building upon the previous stage, we freeze the U-Net backbone part and introduce multiple parallel convolution layers as ensemble learners. These ensemble learners are trained in an individual way that each one has its own loss function for independent backpropagation process. To encourage diverse knowledge acquisition among these learners from the latent representation, we apply different dropout rates to each independent output branch. Moreover, the ConvLSTM module is further finetuned to adapt itself to the changes brought by the learners.

\subsection{Stage 3: Training of Ensemble Controllers}
\begin{figure}[h]
    \centering
    \includegraphics[width=1\linewidth]{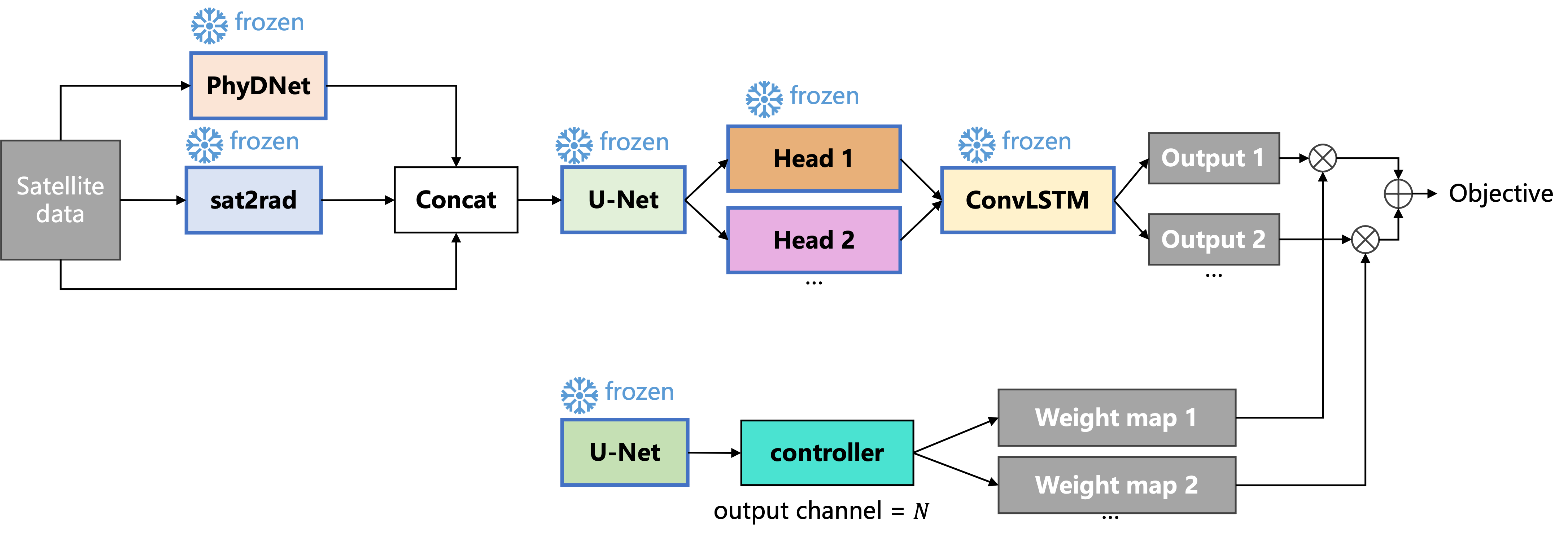}
    \caption{Training of Ensemble Controllers}
    \label{fig:phrase4}
\end{figure}

The corresponding training process is illustrated in Fig~\ref{fig:phrase4}. 
The ensemble controller is responsible for applying learnable weight maps to each output and aggregating the outputs of all the learners to obtain the final result. In this stage, the backbone network remains frozen, and the rainfall probability map obtained from a pretrained U-Net passes through a ConvLSTM module, and its output corresponds to a weight map that is multiplied by the output of each learner. The weighted sum of the outputs is further used to optimized the parameters of the controller.

\section{Experiments}
We conducted all experiments using the training set and validation set of the Weather4cast 2023 Competition dataset. In this section, we experimentally verify our proposed method. The performance evaluation metric of all experiments was selected as mean CSI score for thresholds 0.2, 1, 5, 10, and 15, which rewards correct prediction over the whole gamut of rainfall intensities.

\subsection{Experimental Setting}
\textbf{Weather4cast 2023 Competition Dataset} consist of satellite data and radar ground truth data collected from 7 regions in Europe in 2019 and 2020. The satellite data was obtained from a geostationary meteorological satellite operated by the European Organization for the Exploitation of Meteorological Satellites (EUMETSAT). The radar data is obtained from the Operational Program for Exchange of Weather Radar Information (OPERA) radar network. The satellite data consists of 11 visible, infrared, and water vapor bands modalities, covers a spatial resolution of 12 km $\times$ 12 km with 15-minute intervals. The radar image consists of 252 $\times$ 252 pixels, the same as the satellite data, but covers only a central region of the satellite image with a spatial resolution of 2 km $\times$ 2 km.

\textbf{Implementation detail} To train our framework, the batch size of one GPU is set to 16/8/8 for training stage-1/2/3, respectively, and FP32 training was used. In addition, 20/20/5 epochs were used for the 3 stages and MSE loss is used for all stages. The initial learning rate was set to 1e-3, weight decay to 0.1. AdamW was used as the optimizer. All experiments were performed on Nvidia P100 8 GPUs.

\subsection{Results}

We evaluate the performance of the proposed method on the validation dataset of Weather4cast 2023. The results of baseline model (WeatherFusionNet, short for WFN) as well as our proposed method with different number of learners are compared. For the core test (8h precipitation prediction) setting in Table~\ref{tab:table1}, we can see that ConvLSTM module indeed brings improvement compared with the baseline model, and it is clear that using more learners can improve the performance. Moreover, using multi-head learners achieves better results with high precipitation cases. 

\begin{table}[h]
\center
\caption{8h precipitation prediction results on the Weather4cast 2023 validation dataset. CSI scores of thresholds 0.2, 1, 5, 10, and 15 are used. }
\begin{tabular}{ccccc}
  \toprule
  Threshold & WFN  & 1 learner & 3 learners & 5 learners  \\
  \midrule
  0.2  & 0.31692  & 0.31799  &  0.33249  & 0.34631  \\
  1    & 0.08771  & 0.08940  &  0.14102  & 0.16571  \\
  5    & 0.00190  & 0.00313  &  0.02133  & 0.03821  \\
  10   & 0.0      & 0.00090  &  0.00452  & 0.02691  \\
  15   & 0.0      & 0.0      &  0.00202  & 0.02252  \\
  \midrule
  mean & 0.081306 & 0.08228  &  0.10028  & 0.11993  \\
  \bottomrule
\end{tabular}
\label{tab:table1}
\end{table}

As for the nowcasting (4h precipitation prediction) setting in Table~\ref{tab:table2}, similar conclusions are obtained, while the advantage of using more learners is no longer obvious (3 learners vs 5 learners). 

\begin{table}[h]
\center
\caption{Nowcasting (4h) precipitation prediction results on the Weather4cast 2023 validation dataset. CSI scores of thresholds 0.2, 1, 5, 10, and 15 are used. }
\begin{tabular}{ccccc}
  \toprule
  Threshold & WFN  & 1 learner & 3 learners & 5 learners  \\
  \midrule
  0.2  & 0.36467  & 0.36844  &  0.38724  & 0.38987  \\
  1    & 0.14185  & 0.14616  &  0.20307  & 0.20521  \\
  5    & 0.00356  & 0.00591  &  0.02452  & 0.03734  \\
  10   & 0.0      & 0.00163  &  0.00812  & 0.00865  \\
  15   & 0.0      & 0.0      &  0.00347  & 0.00644  \\
  \midrule
  mean & 0.10201 & 0.10443  &  0.125284  & 0.12950  \\
  \bottomrule
\end{tabular}
\label{tab:table2}
\end{table}

\begin{figure}[t]
    \centering
    \includegraphics[width=1\linewidth]{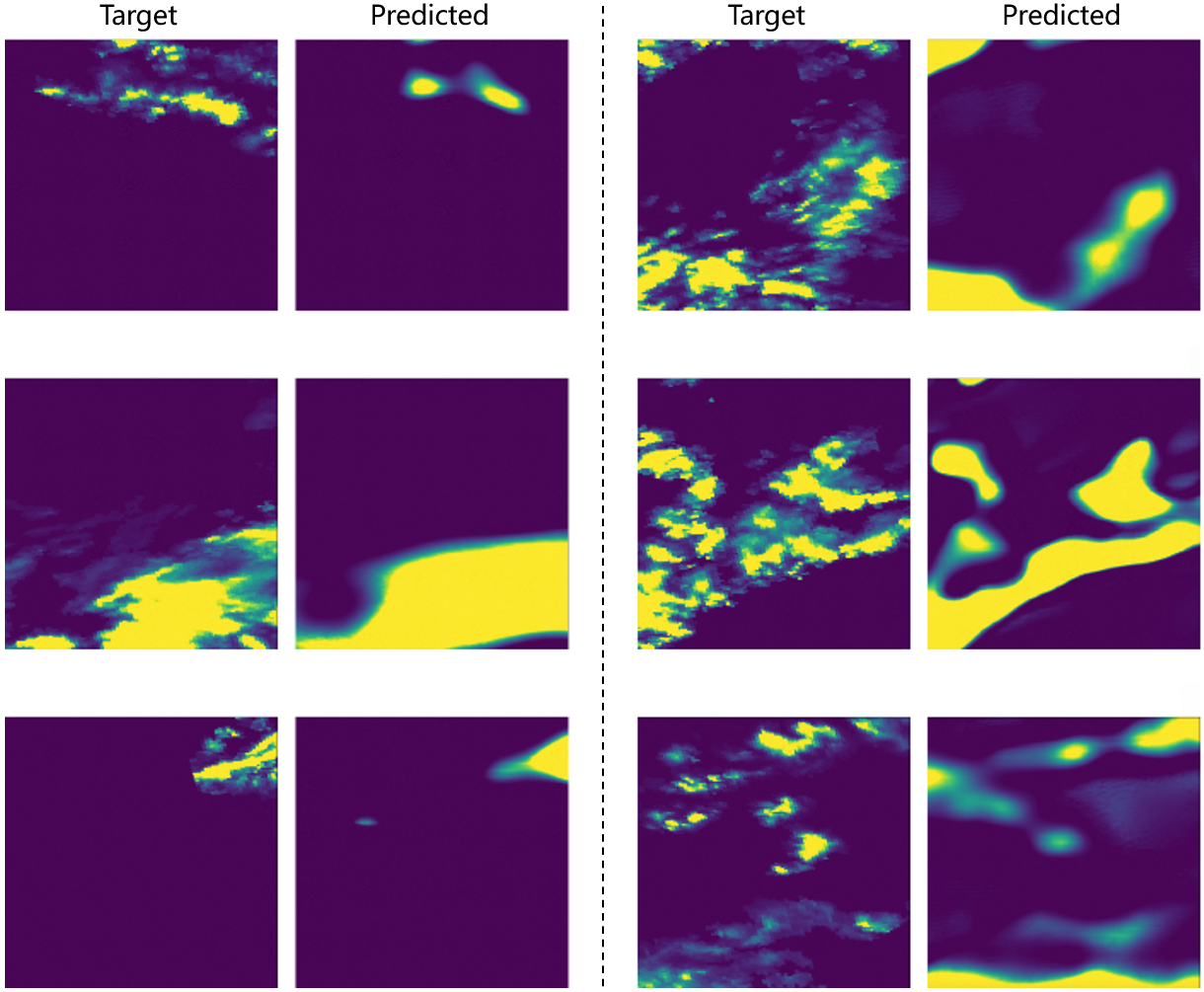}
    \caption{Visualization for some predictions. The left part of each pair is the ground truth image and the right part is the predction.}
    \label{fig:vis}
\end{figure}
We also visualize predictions produced by the proposed method, as shown in Fig~\ref{fig:vis}.  
It is clear that the predictions can locate (high) rainfall regions precisely. However, it is still difficult for it to model complex precipitation distributions. We will try to improve the framework and  solve this problem in our future work.

\section{Conclusions}
In this paper, we propose an ensemble learning framework that leverages multiple
learners for the precipitation prediction task. 
Specifically, multiple learners allow the framework to effectively capture the diverse patterns of precipitation distribution. We further introduce a controller model to generate weight maps for learners to make more reliable decisions. 
The experimental results demonstrate that the proposed method has led to a substantial improvement in scenarios with high precipitation levels. The proposed method achieved 1st place on the core test as well as the nowcasting leaderboards of the Weather4Cast 2023 competition.








\end{document}